%% file: main.tex
\documentclass[runningheads]{llncs}

\usepackage{amsthm}
\usepackage{graphicx}
\usepackage{cite}
\usepackage{amsmath}
\usepackage{amsthm}
\usepackage{booktabs}
\usepackage{algorithmic}
\usepackage{subfigure}

\usepackage{multirow}
\usepackage{color}
\usepackage{bm}

\theoremstyle{definition}
\usepackage{multirow}
\usepackage{amsopn}
\usepackage{makecell}
\usepackage{booktabs}
\usepackage{amsfonts}
\usepackage[ruled,linesnumbered]{algorithm2e}
\usepackage{enumitem}
\usepackage{balance}
\usepackage{bbding}
\usepackage[marginal]{footmisc}
\makeatletter
\newcommand{\printfnsymbol}[1]{%
  \textsuperscript{\@fnsymbol{#1}}%
}
\makeatother
\begin{document}
\author{Xiangyu Zhang \thanks{indicates equal contributions.} \inst{1} \and Zongqiang Kuang \printfnsymbol{1}\inst{1}\and Zehao Zhang\inst{1,2} \and Fan Huang \thanks{indicates corresponding author.} \inst{1} \and Xianfeng Tan\inst{3}}
\authorrunning{Zhang et al.}
\institute{Tencent, Shenzhen, China \\
\email{\{altairzhang,devinkuang,sinohuang\}@tencent.com} \\
\and Tsinghua University, Beijing, China \\
\email{zhangzeh20@mails.tsinghua.edu.cn}
\and Tencent, Beijing, China \\
\email{victan@tencent.com}}

\title{Cold \& Warm Net: Addressing Cold-Start Users in Recommender Systems}

\titlerunning{Cold \& Warm Net}
%
%
\maketitle              
%
\begin{abstract}
Cold-start recommendation is one of the major challenges faced by recommender systems (RS). Herein, we focus on the user cold-start problem. Recently, methods utilizing side information or meta-learning have been used to model cold-start users. However, it is difficult to deploy these methods to industrial RS. There has not been much research that pays attention to the user cold-start problem in the matching stage. In this paper, we propose Cold \& Warm Net based on expert models who are responsible for modeling cold-start and warm-up users respectively. A gate network is applied to incorporate the results from two experts. Furthermore, dynamic knowledge distillation acting as a teacher selector is introduced to assist experts in better learning user representation. With comprehensive mutual information, features highly relevant to user behavior are selected for the bias net which explicitly models user behavior bias. Finally, we evaluate our Cold \& Warm Net on public datasets in comparison to models commonly applied in the matching stage and it outperforms other models on all user types. The proposed model has also been deployed on an industrial short video platform and achieves a significant increase in app dwell time and user retention rate.

\keywords{recommender systems \and cold-start \and Cold \& Warm Net.}
\end{abstract}
\input{introduction}

\input{relatedwork}

\input{method}
\input{experiment}

\input{conclusion}
%
%
%
\bibliographystyle{splncs04}
%
\bibliography{TEX/main}
\end{document}

%% file: introduction.tex
\section{Introduction}
As online information in social media and e-commerce platforms grows explosively, large-scale recommender systems (RS) \cite{linden2003amazon} play an important role in solving the problem of information overload for users. Industrial RS \cite{lv2019sdm} typically contains two stages: matching and ranking. In the matching stage, thousands of items potentially relevant to the user's interests are retrieved from a large-scale candidate pool, which is required to quickly find as many items that satisfy the user's interests as possible. after that, the ranking stage is performed to precisely predict the probability of a user interacting with an item. 

Recently, the matching stage in recommenders \cite{covington2016deep} has been paid increasing attention. Various methods have been applied in the matching stage. Conventional collaborative filtering (CF) method \cite{linden2003amazon} depends on the similarity of interacted items between users for the recommendation. State-of-the-art methods based on reinforcement learning \cite{wang2020kerl}, graph network \cite{ying2018graph} and Multi-Interest network \cite{chai2022user} focus on user behavior sequence representation owned solely by active users with much interaction behavior. However, these models fail to learn high-quality embeddings for the cold-start users with sparse interaction behavior.

Faced with the cold-start problem, side information \cite{zhang2020joint} has been used to provide a better recommendation. However, methods utilizing side information can only benefit part of the users. There are some attempts \cite{dong2020mamo} to introduce meta-learning into recommender systems, which requires the computation of second-order gradients. Therefore, it cannot meet the scalability required by the matching stage of real-world recommendation scenarios. Scalability is the ability to process large-scale information efficiently.

In this paper, modeling cold-start users in the matching stage is our purpose. The core mission of modeling cold-start users is to learn collaborative information between old and cold-start users and train models effectively. Herein, we propose an implicit embedding net based on cold-start and warm-up experts which solves the problems mentioned above efficiently. According to the frequency of interaction, users can be briefly divided into three categories: cold-start users, warm-up users, and active users. The category of users is dynamically changing with the accumulation of interests and behavior, so it is not suitable to use the same strategies for different types of users. Our embedding net based on cold-start and warm-up experts models the dynamic process of cold-start users towards warm-up and active users without compulsory strategies. Overall, the main contributions of this work can be concluded as follows.
\begin{itemize}
\item Dynamic handling of samples. With the division of cold-start and warm-up experts, our Cold \& Warm Net can dynamically represent the users’ interest in cold-start and warm-up phases. Through the gate network, the net can automatically incorporate the results from two experts according to user type and user state. Cold-start and warm-up experts can learn the differences between samples.
\end{itemize}

\begin{itemize}
\item Flexible teacher selector. Dynamic knowledge distillation is applied to cold-start and warm-up experts using a teacher selector. The selector chooses the right teacher for the cold-start expert according to prediction accuracy. By applying dynamic knowledge distillation, it avoids the underfitting of the cold-start expert while preventing the assimilation of two experts after training, which enables learning sufficient information from cold-start users.
\end{itemize}

\begin{itemize}
\item Explicit modeling of behavior bias. Using a bias net to model the behavior bias of cold-start users explicitly. By utilizing mutual information, user features highly relevant to user behavior are selected. With the combination of the similarity score from the original net and the bias score from the bias net, information hidden behind user behavior is thoroughly considered.  
\end{itemize}

%% file: relatedwork.tex
\section{RELATED WORK}\label{sec:related work}

In this section, we review the two-tower models based on embedding which are applied in the matching stage and models targeting the cold-start problem.

\subsection{Two-tower models in the matching stage}\label{subsec:Two-tower models in the matching stage}

One of the challenges faced by RS in the matching stage is that the representations of users and items are not in the same latent space. Models based on embedding learn how to map the sparse user and item vectors in high-dimensional space into dense vectors in low-dimensional space and calculate the inner product or cosine similarity between user and item vectors to obtain a relevance score. The idea of deep learning has been applied in the two-tower models. DSSM \cite{huang2013learning} is a well-known two-tower model utilizing two deep neural networks that map queries and documents into a common space to achieve better search satisfaction. YouTube \cite{covington2016deep} proposes a deep candidate generation model which can effectively learn the embedding of user and item features. \cite{chai2022user} exploits both user profile and behavior information for candidate matching. To tackle sample bias in the matching stage. \cite{huang2020embedding} uses random sampling to acquire negative samples, which successfully bridges the gap in data distribution between training and testing. However, these two-tower models all require close user-item interaction and are incompetent to model cold-start users with rare interaction behavior.

\subsection{Cold-start problem}\label{subsec:Cold-start problem}

The cold-start problem has been one of the long-standing challenges faced by RS. Traditional methods rely on side information to alleviate the cold-start problem, e.g. utilizing social networks among users\cite{zhang2020joint}. Transfer learning-based method \cite{hu2018conet} is also used to deal with the cold-start problem. \cite{dong2020mamo} uses meta-learner to generate cold-start user embedding. However, most of the existing works focus on the cold-start problem in the ranking stage. \cite{zhang2020joint} applies an attention mechanism in multi-channel matching to extract useful feature interactions. \cite{chen2022generative} uses an extra adversarial network to generate cold-start item embedding. \cite{hao2021pre} simulates the cold-start scenarios from the users/items with sufficient interactions and takes the embedding reconstruction as the pretext task. To our knowledge, we are the first to propose an embedding net that dynamically models different types of users in the matching stage.

%% file: method.tex
\section{METHOD}\label{sec:method}

\subsection{Problem description}
The objective of the matching stage for RS is to retrieve Top $K$ relevant items from a large-scale candidate pool $\mathcal{I}$ for each user $u \in \mathcal{U} $. To achieve this target, a matching model is built. The input of model is a tuple $(\mathcal{\mathcal{X}}_u, \mathcal{X}_i)$, where $\mathcal{X}_u$ denotes user features and $\mathcal{X}_i$ denotes item features.
Modeling cold-start users is tough due to the lack of behavior. The core task of Cold \& Warm Net is to learn a function that can map original features into user representations. The function can be formulated as:
\begin{equation}
{\overrightarrow{\boldsymbol{e}}_u}=f_{user}\left({\mathcal{X}_u}\right)
\end{equation}
where $\overrightarrow{\boldsymbol{e}}_u \in \mathbb{R}^{1 \times d}$ denotes the representation vector of user $u$, $d$ the dimensionality. In addition, the representation vector of target item $i$ is
obtained by a function:
\begin{equation}
{\overrightarrow{\boldsymbol{e}}_i}=f_{item}\left({\mathcal{X}_i}\right)
\end{equation}
where $\overrightarrow{\boldsymbol{e}}_i \in \mathbb{R}^{1 \times d}$ denotes the representation vector of item $i$. Finally, The top $K$ relevant items are retrieved according to the scoring function:
\begin{equation}
f_{score}\left({\overrightarrow{\boldsymbol{e}}_u, \overrightarrow{\boldsymbol{e}}_i}\right)={\overrightarrow{\boldsymbol{e}}_u}\cdot{\overrightarrow{\boldsymbol{e}}_i}
\end{equation}

\subsection{Cold \& Warm Net}

\begin{figure*}[t!]
    \centering
    \includegraphics[width=\textwidth]{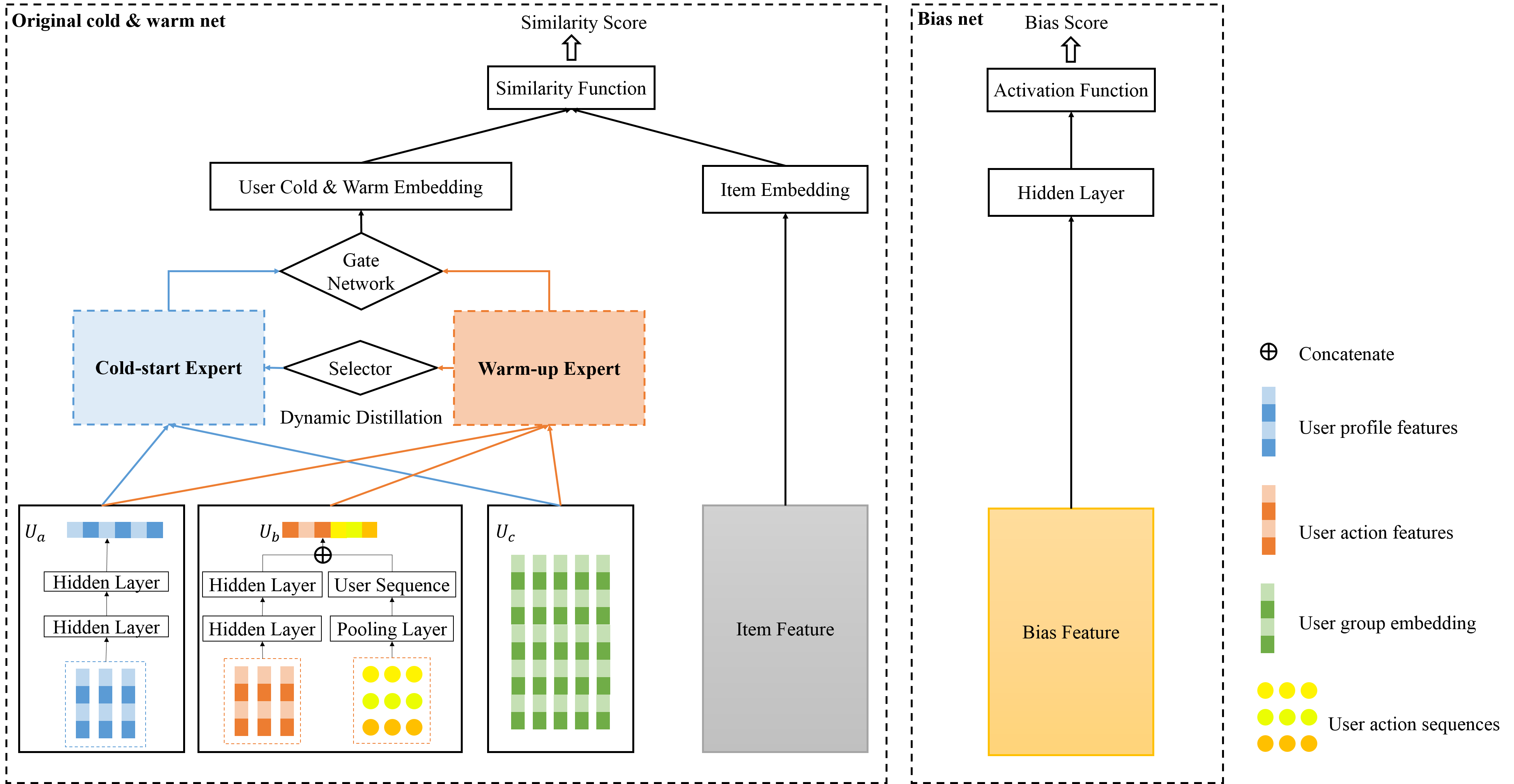}
    \caption{Cold \& Warm Net.}
    \label{fig2}
\end{figure*}

  As shown in Figure \ref{fig2}, our Cold \& Warm Net consists of two subnets: original cold \& warm net and bias net. The original cold \& warm net uses user features $\mathcal{X}_u$ and item features $\mathcal{X}_i$ as input while the bias net takes bias features $\mathcal{X}_b$ as input. We divide user features into two categories: user profile features $\mathcal{X}_{up}$(e.g., gender and age) and user action features $\mathcal{X}_{ua}$(also called user behavior). Taking different user features as input, we attain output embedding: user profile embedding $\overrightarrow{\boldsymbol{e}}_{up} \in\mathbb{R}^{1 \times d}$ and user action embedding $\overrightarrow{\boldsymbol{e}}_{ua}\in \mathbb{R}^{1 \times d}$. Besides, user group embedding $E_{ug}  \in \mathbb{R}^{m\times d}$ is provided as prior information for all users. Firstly, We use a pre-trained model for getting all active-user embeddings. Then, taking all active-user embeddings as input, the k-means algorithm is used to attain $m$ clusters. Finally, we use average pooling to aggregate the embeddings of each cluster for generating $E_{ug}$.  The three parts are defined as $U_a$, $U_b$ and $U_c$, which are fed into user cold \& warm embedding layer to generate user embedding $\overrightarrow{\boldsymbol{e}}_u$. Along with item embedding $\overrightarrow{\boldsymbol{e}}_i$, similarity score $y_{sim\_score}$ is calculated as follows.

\begin{equation}\label{eq:score}
y_{sim\_score}=\frac{\overrightarrow{\boldsymbol{e}}_u\cdot{\overrightarrow{\boldsymbol{e}}_i}}{\|\overrightarrow{\boldsymbol{e}}_u\|\|\overrightarrow{\boldsymbol{e}}_i\|}
\end{equation}
The similarity score $y_{sim\_score}$ from the original cold \& warm net and the bias score $y_{bias\_score}$ from the bias net constitute our final output:
\begin{equation}\label{eq:score}
y=sigmoid(y_{sim\_score}+y_{bias\_score})
\end{equation}

\subsubsection{User cold \& warm embedding layer}

\begin{figure*}[t!]
    \centering
    \includegraphics[width=\textwidth]{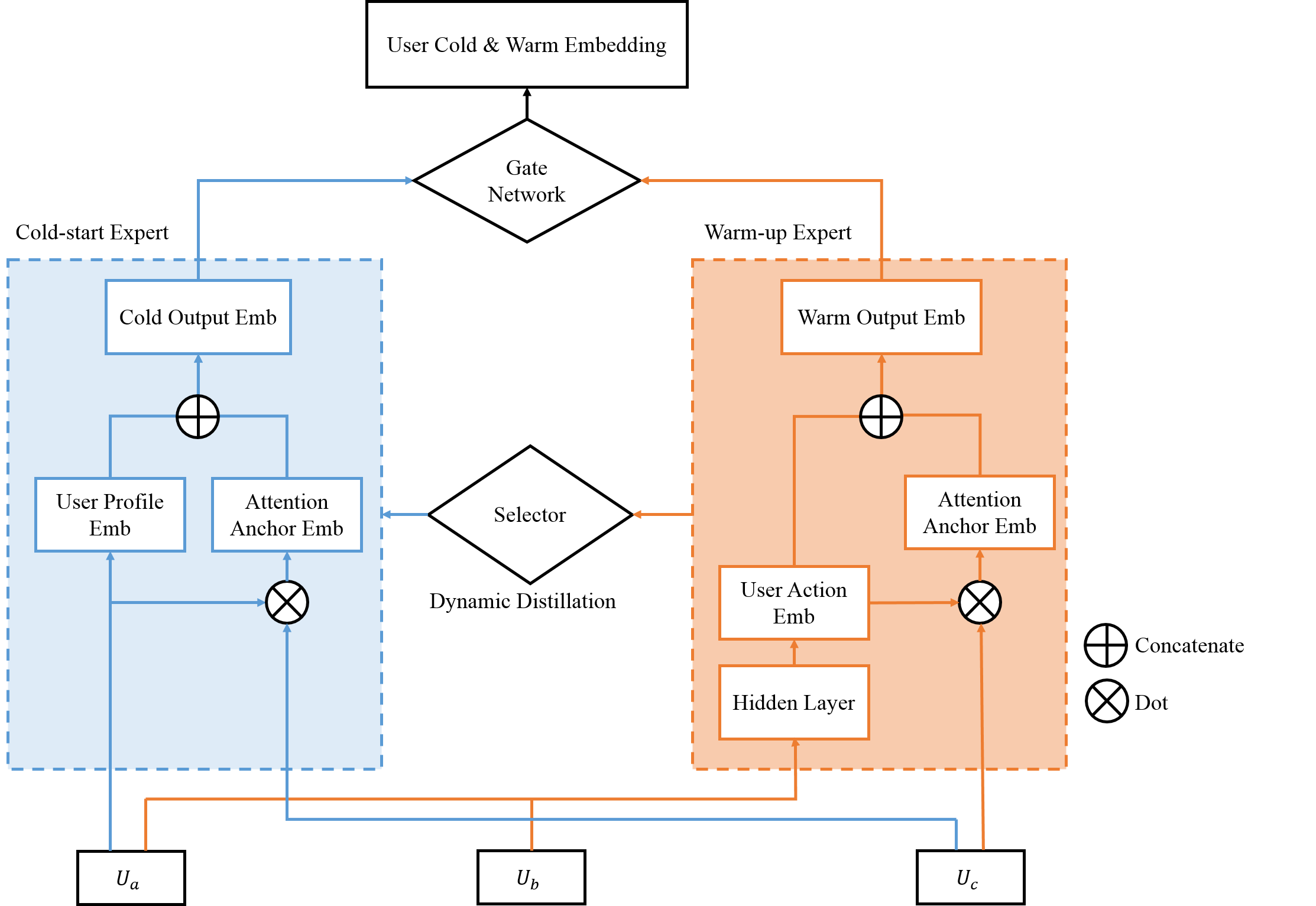}
    \caption{User cold \& warm embedding layer.}
    \label{fig3}
\end{figure*}

As shown in Figure \ref{fig3}, our user cold \& warm embedding layer is mainly composed of two experts: the cold-start expert and the warm-up expert. To extract and learn valid information that matches the current user, we use the attention mechanism to retrieve the prior information from $E_{ug}$  which contains all user group pre-trained embedding. Cold-start expert takes input from $U_a$ and $U_c$ which contain user profile information and user group information. Attention embedding takes prior user group information to assist in modeling cold-start users, which can be formulated as:
\begin{equation}
\overrightarrow{\boldsymbol{e}}_{cold}^a=softmax(\frac{\overrightarrow{\boldsymbol{e}}_{up} E_{ug}^T}{\sqrt{d}})E_{ug}
\end{equation}
where $\overrightarrow{\boldsymbol{e}}_{cold}^a \in \mathbb{R}^{1 \times d}$ means pre-trained embedding retrieved from $E_{ug}$ using attention mechanism \cite{kang2018self}.
The output embedding of cold-start expert $\overrightarrow{\boldsymbol{e}}_{cold}$ is:
\begin{equation}
\overrightarrow{\boldsymbol{e}}_{cold}=mlp(\overrightarrow{\boldsymbol{e}}_{up}; \overrightarrow{\boldsymbol{e}}_{cold}^a)
\end{equation}
Where $\overrightarrow{\boldsymbol{e}}_{cold} \in \mathbb{R}^{1 \times d}$. The warm-up expert takes input from $U_a$, $U_b$ and $U_c$. It is designed for users who possess user profile features $\mathcal{X}_{up}$ and user action features $\mathcal{X}_{ua}$. Taking $\mathcal{X}_{up}$ and $\mathcal{X}_{ua}$ as input, we attain embedding $\overrightarrow{\boldsymbol{e}}_{ut} \in \mathbb{R}^{1 \times d}$. Through the assistance of attention anchor embedding, the output embedding of warm-up expert $\overrightarrow{\boldsymbol{e}}_{warm}$ is defined as follows.
\begin{equation}
\overrightarrow{\boldsymbol{e}}_{warm}^a=softmax(\frac{\overrightarrow{\boldsymbol{e}}_{ut} E_{ug}^T}{\sqrt{d}}) E_{ug}
\end{equation}
\begin{equation}
\overrightarrow{\boldsymbol{e}}_{warm}=mlp(\overrightarrow{\boldsymbol{e}}_{ut}; \overrightarrow{\boldsymbol{e}}_{warm}^a)
\end{equation}
where $\overrightarrow{\boldsymbol{e}}_{warm} \in \mathbb{R}^{1 \times d}$. A gate network is used to produce weights for experts.
\begin{equation}
w_{cold}, w_{warm}=f_{weight}(\mathcal{X}_{us})
\end{equation}
where $\mathcal{X}_{us}$(such as login state and active degree) denotes state feature. The output user cold \& warm embedding is:
\begin{equation}
\overrightarrow{\boldsymbol{e}}_u={w_{cold}} \cdot \overrightarrow{\boldsymbol{e}}_{cold}+ {w_{warm}} \cdot \overrightarrow{\boldsymbol{e}}_{warm}
\end{equation}
The above expression can be understood as a weighted summation of cold-start expert and warm-up expert.

\subsubsection{Dynamic knowledge distillation}

With a mix of experts, cold-start expert suffers from underfitting due to limited information from cold-start users. The reason is that the warm-up expert learns better for active users which own rich behavior features. To avoid underfitting of the cold-start expert, we invent dynamic knowledge distillation(DKD) to distill information from the warm-up expert to the cold-start expert. Binary cross entropy is selected as our loss function. The major loss function $L$:
\begin{equation}
L=-\frac{1}{N} \sum_{i=1}^{N}\left(y_{i} \log \hat{y}_{i}+\left(1-y_{i}\right) \log \left(1-\hat{y}_{i}\right)\right)
\end{equation}
 where $N$ denotes the number of samples, $y_i$ is the label for each sample. $\hat{y}_{i}$ denotes the predicted result. $L$ is the loss function except for the cold-start expert. Besides, The auxiliary loss function $L_d$ 
from dynamic knowledge distillation:
\begin{equation}
L_{d}=-\frac{1}{N} \sum_{i=1}^{N} l_{d}
\end{equation}
where $l_{d}$ is the loss function of DKD for each sample, which is shown in Algorithm \ref{alg:A}. For each sample, we compare the cross entropy loss of cold-start expert $l(\hat {y_i}^c, y_i)$ and warm-up expert $l(\hat {y_i}^w, y_i)$. If $l(\hat {y_i}^c, y_i)\le l(\hat {y_i}^w, y_i)$, it indicates that output from the cold-start expert taught by the label is better and there is no need to distill information from the warm-up expert. Otherwise, the cold-start expert taught by the label is not enough and it is necessary to learn from the warm-up expert. $\hat {y_i}^c$ denotes the predicted label for the cold-start expert and $\hat {y_i}^w$ is the predicted label for the warm-up expert. The teacher for knowledge distillation is dynamic to make sure the cold-start expert could learn effective information from the teacher. The loss function of the cold-start expert is defined as $L_o$.
\begin{equation}
L_o=L+\alpha \cdot L_d
\end{equation}
where $\alpha$ is hyperparameter. $\alpha$ determines the strength of distillation from the warm-up expert.
\linespread{1.2}
\begin{algorithm}[t]
	\SetKwInOut{Input}{Input}
	\SetKwInOut{Output}{Output}
	\BlankLine
	
	\ForEach{sample from batch samples}{
    Calculate $l(\hat {y_i}^c, y_i)$, $l(\hat {y_i}^w, y_i)$
    
     \eIf  {$l(\hat {y_i}^w, y_i) > l(\hat {y_i}^c, y_i)$}    
     {$l_d = l(\hat {y_i}^c, \hat {y_i}^w) $\;}
     {$l_d = 0$\;}

	}
	
	\caption{Dynamic knowledge distillation.\label{alg:A}}
\end{algorithm}
\linespread{1}

\subsubsection{Bias net.}
To solve the behavior bias when modeling cold-start users, an additional bias net is applied. The reason why bias net is effective is that the behavior bias is large between cold-start users and active users in real-world recommendation scenarios. For example, active users have several times more click rates than new users, so a bias net is indispensable for describing the bias. We aim to find a set of user features $\mathcal{X}_b$ which are highly relevant to user behavior and feed them into the bias net to get $y_{bias\_score}$. To mine the key features of the behavior bias, mutual information is used to measure the relevance between user features and behaviors. We select the top $\beta$ relevant features as our bias features $\mathcal{X}_b$. The output bias score $y_{bias\_score}$ is:
\begin{equation}
y_{bias\_score} = f_{bias}(\mathcal{X}_b)
\end{equation}
$y_{bias\_score}$ is used to characterize the bias of target behaviors of people.

%% file: experiment.tex
\section{Experiments}\label{sec:experiment}
\subsection{Offline Evaluation}\label{subsec:Offline Evaluation}
In this section, we compare our Cold \& Warm Net with existing methods applied in the matching stage in terms of recommendation accuracy on two datasets. 
\subsubsection{Datasets and experimental setup}\label{subsec:Datasets and experimental setup}
Two datasets are chosen to evaluate the recommendation performance in the matching stage. One is MovieLens 1M\footnote{https://grouplens.org/datasets/movielens/1m/}, which is one of the most common datasets used for recommendations. Also, we collect a real-world large-scale dataset from the Little-World\footnote{Little-World is a short video platform in QQ,  which allows users to create and share micro-videos. Note that we anonymize the data and conduct strict desensitization processing. The data may be made public in the future.}. The statistics of datasets are shown in Table \ref{tab:dataset}. Hit rate (HR) and Normalized discounted cumulative gain (NDCG) are adopted as the main metric to evaluate the performance of models in the matching stage, define as:
\begin{equation}
HitRate@K = \sum_{(u,i)\in {T}} \frac{I(target items occu in topK)} {\lvert {T} \rvert}
\end{equation}

\begin{equation}
NDCG@K = \frac{1}{|U|} \sum_{u \in U} \frac{DCG_{k}^{u}}{IDCG_{k}^{u}}
\end{equation}

\begin{equation}
DCG_{k}^{u}=\sum_{r=1}^{k} \frac{2^{R_{ur}}-1}{\log _{2}(1+r)}
\end{equation}
where ${T}$ denotes the test set containing pair of user and item and $I$ denotes the indicator function. $R_{ur}$, $U$, and $IDCG_{k}^{u}$ are the real rating of user $u$ for the $r$-th ranked item, a set of users in the test data and the best possible $DCG_{k}^{u}$ for 
user $u$, respectively. Specially, in the matching stage, $R_{ir} \in \left \{0, 1 \right \}$.

\begin{table}[t]
\caption{Statistics of the datasets.}\label{tab:dataset}
\begin{center}
\begin{tabular}{cccc}
\toprule
Dataset & \# User & \# Items & \# Interaction \\ 
\midrule 
MovieLens 1M	& 6040	& 3706	&1,000,209\\
Little-World	& 433,549	&406,140	&15,200,286\\
\bottomrule
\end{tabular}
\label{table1}
\end{center}
\end{table}

\subsubsection{Comparing methods}
The following methods widely applied in the matching stage in industry RS are used to compare with our Cold \& Warm Net. 
\begin{itemize}
\item FM\cite{rendle2010factorization} A model that utilizes the feature vectors of query and item and feeds them into FM layer.
\end{itemize}
\begin{itemize}
\item YouTubeDNN\cite{covington2016deep} One of the most commonly used models in the recommendation industry which applies deep neural network to generate item and user embedding.
\end{itemize}
\begin{itemize}
\item DSSM\cite{huang2013learning} A popular model applied in the matching 
stage which makes use of rich content features of user and item.
\end{itemize}
\begin{itemize}
\item Mind\cite{li2019multi} the first attempt in representing a user with multiple interest vectors via deep neural network structures.
\end{itemize}
\begin{itemize}
\item UMI\cite{chai2022user} State-of-the-art model that relies on multiple user interest representations to achieve superior recommendation accuracy.
\end{itemize}
The above models are implemented by Tensorflow and Faiss is used to retrieve the top $K$ items from the item pool. The embedding dimension and batch size are set to 32 and 256 respectively for all models. To ensure a fair comparison, for each model, hyperparameters are tuned to achieve the best performance. For Cold \& Warm Net, hyperparameters $\alpha$, $\beta$ are set to 5e-2 and 10 respectively.
\begin{table*}[t]
\caption{Performance comparison of different models in terms of HR and NDCG}\label{tab:results}
\begin{center}
\resizebox{\linewidth}{!}{
\begin{tabular}{|c|c|c|c|c|c|c|c|c|}
 \multicolumn{9}{c}{(a)Results on full users} \\
 \hline
\multirow{2}{*}{Models} &
\multicolumn{4}{c|}{MovieLens 1M} &
\multicolumn{4}{c|}{Little-World} \\
\cline{2-9}
 &HR@50	& HR@100	& NDCG@10	& NDCG@50   & HR@50	& HR@100	& NDCG@10	& NDCG@50 \\
\hline
FM	& 0.0969	& 0.1922	& 0.0099	& 0.0262	& 0.0513	&0.0754	& 0.0100	& 0.0173 \\			
\hline
YouTubeDNN&	0.1399&	0.2548&	0.0153&	0.0378&	0.0862&	0.1461&	0.0110&	0.0245 \\
\hline
DSSM&	0.2013&	0.3151&	0.0226&	0.0520&	0.0913&	0.1511&	0.0116&	0.0260\\			
\hline
Mind&	0.2019&	0.3322&	0.0238&	0.0612&	0.0917&	0.1530&	0.0118&	0.0262\\
\hline
UMI&	0.2348*&	0.3697*&	0.0305*	&0.0664*&	0.0920*&	0.1546*&	0.0119* &	0.0270*\\
\hline
Cold \& Warm&	$\bm {0.2556}$&	$\bm{0.3932}$&	$\bm{0.0369}$&	$\bm{0.0750 }$& $\bm {0.1122}$ &$\bm {0.1792}$	&$\bm{0.0155}$	&$\bm {0.0325}$ \\
\hline
\%improve.&	8.86\%&	6.35\%&	20.98\%&	12.95\%&	21.95\%&	15.91\%&	30.25\%&	20.37\% \\
\hline
\end{tabular}
}
\ \notag\ \\
\resizebox{\linewidth}{!}{
\begin{tabular}{|c|c|c|c|c|c|c|c|c|}
 \multicolumn{9}{c}{(b)Results on cold-start users} \\
 \hline
\multirow{2}{*}{Models} &
\multicolumn{4}{c|}{MovieLens 1M} &
\multicolumn{4}{c|}{Little-World} \\
\cline{2-9}
 &HR@50	& HR@100	& NDCG@10	& NDCG@50   & HR@50	& HR@100	& NDCG@10	& NDCG@50 \\
\hline
FM	&0.1568	&0.2953	&0.0211	&0.0461&	0.0710&	0.1047&	0.0147&	0.0242\\
\hline
YouTubeDNN&	0.2444&	0.3849&	0.0236&	0.0639&	0.1088&	0.1768&	0.0138&		0.0311\\
\hline
DSSM&	0.3666*&	0.5356*&	0.0657*&	0.1173*&	0.1109*&	0.1775*&	0.0159*&	0.0326*\\			
\hline
Mind&	0.3259&	0.4807&	0.0485&	0.1008&	0.1074&	0.1771&	0.0132& 0.0300\\
\hline
UMI&	0.3360&	0.4705&	0.0493&	0.1002&	0.1103&	0.1671&	0.0143& 0.0281\\
\hline
Cold \& Warm&	$\bm {0.4094}$&	$\bm{0.5866 }$&	$\bm{0.0678}$&	$\bm{0.1265 }$& $\bm {0.1435}$ &$\bm{0.2200 }$	&$\bm{0.0215}$	&$\bm{0.0418 }$ \\				
\hline
\%improve.&	11.67\%&	9.52\%&	3.19\%&	7.84\%&	29.39\%&	23.94\%&	35.22\%&	28.22\% \\
\hline
\end{tabular}
}
\label{table2}
\end{center}
\end{table*}
\subsubsection{Experimental results}
Table \ref{tab:results} summarizes the performance of Cold \& Warm Net in comparison with different models applied in the matching stage in terms of HR@$K$($K$=50, 100) and NDCG@$K$($K$=10, 50). Obviously, Cold \& Warm Net achieves the best recommendation performance among all models on different user categories. FM performs worst among all models revealing the power of deep learning. UMI and Mind which utilize multiple interest representations of user generally performs better than YouTubeDNN which only uses single-user interest representation. UMI performs better than Mind due to exploiting both user profile and behavior information for candidate matching. However, Mind and UMI perform worse than the DSSM model for cold-start users, which may be because cold-start users lack abundant interests. Results on two types of users justify the performance of Cold \& Warm Net on different types of users, effectively solving the user cold-start problem faced in recommender systems.

\begin{table*}[t]
\caption{Ablation study of Cold \& Warm Net.}\label{tab:Ablation study}
\begin{center}
\resizebox{\linewidth}{!}{
\begin{tabular}{|c|c|c|c|c|c|c|c|c|}
 \hline
\multirow{3}{*}{Models} &
\multicolumn{4}{c|}{MovieLens 1M} &
\multicolumn{4}{c|}{Little-World} \\
\cline{2-9}
 & \multicolumn{2}{c|}{Full users} &
\multicolumn{2}{c|}{Cold-start users} &
\multicolumn{2}{c|}{Full users} &
\multicolumn{2}{c|}{Cold-start users} \\
\cline{2-9}
  & HR@100	& NDCG@10 &HR@100	& NDCG@10 &HR@100	& NDCG@10 &HR@100	& NDCG@10	 \\
\hline
Cold \& Warm	& $\bm {0.3932}$	&$\bm {0.0369}$	&$\bm {0.5866}$	&$\bm {0.0678}$ &$\bm {0.1792}$ &$\bm {0.0155}$ &$\bm {0.2200}$ &$\bm {0.0215}$ \\	
\hline
w/o DKD	& 0.3869&	0.0318&	0.5540&	0.0581	& 	0.1703& 0.0144& 0.1987&0.0190 \\		
\hline
w/o Bias Net	&0.3930	&0.0367	&0.5682	&0.0632 &0.1761 &0.0147 &0.2122 &	0.0203 \\	
\hline
\end{tabular}
}
\label{table3}
\end{center}
\end{table*}

Table \ref{tab:Ablation study} summarizes the result of the ablation study. It is conducted to evaluate the contribution of the dynamic knowledge distillation(DKD) module and bias net. DKD and bias net designed for cold-start users contribute mainly to solving the problem of modeling cold-start users. For cold-start users, applying DKD brings an increase of 5.88\% and 10.72\% in HitRate@100 on two datasets while adding bias net brings an increase of 3.24\% and 3.68\% in HitRate@100. The major boost from DKD may be due to the fact that the cold-start expert learns better user representation with the assistance of the warm-up expert. 
\begin{table}[t]
\caption{Influence of Dynamic Knowledge Distillation on weights ($w_{cold}$, $w_{warm}$).}\label{tab: Dynamic Knowledge Distillation on weights}
\begin{center}
\begin{tabular}{|c|c|c|}
 \hline
Metrics & Cold-start expert & Warm-up expert\\
 \hline
Weights (w/o DKD) & 0.0410 & 0.9590\\
 \hline
Weights (DKD) & 0.3140 & 0.6860\\
\hline
\end{tabular}
\label{table4}
\end{center}
\end{table}
\begin{table*}[t]
\caption{Influence of Dynamic Knowledge Distillation on AUC.}\label{tab: Dynamic Knowledge Distillation}
\begin{center}
\resizebox{\linewidth}{!}{
\begin{tabular}{|c|c|c|c|c|c|c|}
 \hline
\multirow{2}{*}{Metrics}  & \multicolumn{3}{c|}{w/o DKD} &
\multicolumn{3}{c|}{DKD}  \\
\cline{2-7}
 & Cold-start expert & Warm-up expert & Whole & Cold-start expert & Warm-up expert & Whole  \\
\cline{2-7}
\hline
full users& 0.5770 & 0.9255 & 0.9267 & $\bm{0.8772}$ & 0.8993  & $\bm{0.9279}$	\\
\hline
cold-start users & 0.5675 & 0.7279 & 0.7281 &$\bm{0.7384}$ & 0.7434 & $\bm{0.7548}$	\\
\hline
\end{tabular}
}
\label{table5}
\end{center}
\end{table*}
\subsection{Analysis of Dynamic Knowledge Distillation}\label{subsec:Analysis of Dynamic Knowledge Distillation}
In this section, the influence of DKD has been analyzed based on the Little-World dataset. AUC is chosen as the evaluation metric. As shown in Table \ref{tab: Dynamic Knowledge Distillation on weights}, by applying dynamic knowledge distillation, it greatly improves the weight of cold-start expert $w_{cold}$ from 0.0410 to 0.3140, which allows the cold-start expert to learn sufficient information either from warm-up expert or label and avoids underfitting of cold-start expert. It can be seen from Table \ref{tab: Dynamic Knowledge Distillation} that after applying DKD, the AUC of the cold-start expert increases obviously, which demonstrates the effect of DKD on enabling sufficient training. The AUC of the warm-up expert decreases on the train set because DKD reduces losses flowing to the warm-up expert. Therefore, the major contribution of AUC comes from sufficient learning of the cold-start expert. Improved AUC of the whole model on the test set also justifies the influence of DKD on cold-start users.
\subsection{Online Experiment}\label{subsec:Online Experiment}
Finally, we deploy Cold \& Warm Net in the real-world recommending scenario of Little-World. User retention rate (URR) and app dwell time (APT) are used as the main metrics for cold-start users. All online experimental results are averaged over a week’s A/B test on Little-World. The result shows that Cold \& Warm Net brings an increase of 3.27\% in APT and 1.01\% in URR for cold-start users. Meanwhile, we compare Mind and Cold \& Warm Net using the DSSM model as the baseline. Video play integrity (VPI) and video skip rate (VSR) are used as the main metrics for user satisfaction evaluation. The result in Table \ref{tab:Online experimental results} shows that Cold \& Warm Net outperforms the DSSM model on both VPI and VSR, which indicates successfully modeling cold-start users and improving user satisfaction.

\begin{table}[t]
\caption{Online experimental results. Cold \& Warm Net performs better in terms of VPI and VSR on cold-start users. DSSM model as the baseline. }\label{tab:Online experimental results}
\begin{center}
\begin{tabular}{|c|c|c|}
 \hline
Models&	VPI&	VSR\\
\hline
Cold \& Warm &	$\bm {+23.34\%}$	&$\bm{-14.30\%}$ \\
\hline
Mind &	-2.05\%	& +2.76\% \\
\hline
\end{tabular}
\label{table6}
\end{center}
\end{table}

%% file: conclusion.tex
\section{Conclusion}
User cold-start problem in the matching stage is a critical challenge faced by RS. However, the solutions are rare both in academia and industry. In this paper, we propose Cold \& Warm Net which effectively solves the problem for cold-start users, while in the meantime satisfying the scalability required by the billion-scale matching stage. We first construct our network with two experts and incorporate a gate network to combine results according to the user state. Bias net and DKD module responsible for modeling cold-start users are incorporated. Finally, we evaluate our model through offline and online experiments and it achieves an obvious increase in recommendation performance.